\documentclass[aps,prl,amsmath,preprint,groupedaddress,showpacs,letterpaper]{revtex4}

\usepackage{graphicx}

\begin{document}
\title{Observation of the Dynamical Casimir Effect in a Superconducting Circuit}

\author{C.M. Wilson$^{1}$,G. Johansson$^{1}$, A. Pourkabirian$^{1}$, J.R. Johansson$^{2}$, T. Duty$^{3}$, F. Nori$^{2,4}$ \& P. Delsing$^1$}


\affiliation{$^1$Depart. of Microtechnology and Nanoscience, Chalmers University of Technology, G\"oteborg, Sweden}
\affiliation{$^2$Advanced Science Institute, RIKEN, Wako-shi, Saitama 351-0198, Japan}
\affiliation{$^3$University of New South Wales, Sydney, NSW, 2052 Australia}
\affiliation{$^4$University of Michigan, Ann Arbor, MI 48109, USA}


\date{\today}
 \maketitle     
One of the most surprising predictions of modern quantum theory is that the vacuum of space is not empty. In fact, quantum theory predicts that it teems with virtual particles flitting in and out of existence. While initially a curiosity, it was quickly realized that these vacuum fluctuations had measurable consequences, for instance producing the Lamb shift\cite{OrvilScully:1997p3464} of atomic spectra and modifying the magnetic moment for the electron\cite{Greiner:2008p3477}. This type of renormalization due to vacuum fluctuations is now central to our understanding of nature. However, these effects provide indirect evidence for the existence of vacuum fluctuations. From early on, it was discussed if it might instead be possible to more directly observe the virtual particles that compose the quantum vacuum.  40 years ago, Moore\cite{Moore:1970} suggested that a mirror undergoing  relativistic motion could convert virtual photons into directly observable real photons.  This effect was later named the dynamical Casimir effect (DCE).  Using a superconducting circuit, we have observed the DCE for the first time.  The circuit consists of a coplanar transmission line with an \textit{electrical length} that can be changed at a few percent of the speed of light. The length is changed by modulating the inductance of a superconducting quantum interference device (SQUID) at high frequencies ($\sim 11$ GHz).  In addition to observing the creation of real photons, we observe two-mode squeezing of the emitted radiation, which is a signature of the quantum character of the generation process.

That a mirror can be used to measure vacuum fluctuations was first predicted by Casimir in 1948\cite{Casimir:1948}.  Casimir predicted that two mirrors, i.e. perfectly conducting metal plates, held parallel to each other in vacuum will experience an attractive force.  Essentially, the mirrors reduce the density of electromagnetic modes between them.  The vacuum radiation pressure between the plates is then less than the pressure outside, generating the force.  As this \textit{static} Casimir effect can then be explained by a mismatch of vacuum modes in space, the \textit{dynamical} Casimir effect can be seen as arising from a mismatch of vacuum modes in \textit{time}.  As the mirror moves, it changes the spatial mode structure of the vacuum.  If the mirror's velocity, $v$, is slow compared to the speed of light, $c$, the electromagnetic (EM) field can adiabatically adapt to the changes and no excitation occurs.  If instead $v/c$ is not negligible, then the field cannot adjust smoothly and can be nonadiabatically excited out of the vacuum.  

The static Casimir effect can also be calculated in terms of the electrical response of the mirrors to the EM field\cite{Lamoreaux:2007}.  A similar complementary explanation exists for the DCE\cite{Moore:1970}.  Theoretically, the ideal mirror represents a boundary condition for the EM field, in particular, that the electric field is zero at the surface.  This boundary condition is enforced by the flow of screening currents in the metal.  A mirror moving in a finite EM field then losses energy as the screening currents will emit EM radiation, as in an antenna.  Classically, we expect this radiation damping to be zero in a field-free region.  In quantum theory, however, the screening currents must always act against the vacuum fluctuations.  Therefore, even moving in the vacuum will cause a mirror to emit real photons in response to the vacuum fluctuations.  

If we consider the literal experiment of moving a physical mirror near the speed light, we quickly see that this experiment is not feasible.  Braggio et al. considered\cite{Braggio:2005p687} the case of moving a typical microwave mirror in an oscillating motion at a frequency of 2 GHz with a displacement of 1 nm.  This produces a velocity ratio of only $v/c \sim 10^{-7}$ with an expected photon production rate of approximately 1 per day. Nevertheless, it requires an input of mechanical power of 100 MW while, at the same time, the system would need to be cooled to $\sim 20$ mK to ensure that the EM field is in its vacuum state.  These difficulties have lead to a number of alternative proposals\cite{YABLONOVITCH:1989p1128,LOZOVIK:1995p2780,Dodonov:1993p2799,Liberato:2007p2806,Gunter:2009p2785,Johansson:2009p2003,Johansson:2010p3255,Wilson:2010p3328,Nation:2011p3391}, for instance utilizing surface acoustic waves, nanomechanical resonators, or by modulating the electrical properties of cavities.  

Here we investigate one such proposal using a superconducting circuit\cite{Johansson:2009p2003,Johansson:2010p3255}: an open transmission line terminated by a SQUID.  A SQUID is comprised of two Josephson junctions connected in parallel to form a loop.  At the frequencies studied here, the SQUID acts as a parametric inductor whose value can be tuned by applying a magnetic flux, $\Phi_{\rm ext}$, through the SQUID loop.  When placed at the end of a transmission line, this SQUID can then be used to change the line's boundary condition.  In previous work, we showed that this tuning can be done on very short time scales\cite{Sandberg:2008p1404,Wilson:2010p3328}.  The changing inductance can be described as a change in the \textit{electrical length} of the transmission line and, in fact, provides the same time-dependent boundary condition as the idealized moving mirror\cite{Fulling:1976p2842,Lambrecht:1996p1645}.  In the same way as for the mirror, the boundary condition is enforced by screening currents that flow through the SQUID.  Unlike the mirror, the effective velocity of the boundary, defined as the rate of change of the electrical length, can be very large, approaching $v/c \sim 0.05$ for a 10\% modulation of the SQUID inductance.  The photon production rate is therefore predicted to be several orders of magnitude larger than in other systems.

Quantum theory allows us to make more detailed predictions than just that photons will simply be produced.  If the boundary is driven sinusoidally at an angular frequency $\omega_d = 2\pi f_d$, then it is predicted that photons will be produced in pairs such that their frequencies, $\omega_1$ and $\omega_2$, sum to the drive frequency, i.e., we expect $\omega_d = \omega_1 + \omega_2$.  This pairwise production implies that the EM field at these frequencies, symmetric around $\omega_d/2$, should be correlated.  In detail, we can predict that the field should exhibit what is known as two-mode squeezing\cite{CAVES:1985p2983}.  These correlations are a signature of the two-photon nature of the photon generation process.

Theoretically, we treat the problem as a scattering problem in the context of quantum network theory\cite{Yurke:1984p776}.  For superconducting circuits, it is convenient to describe the EM field in the transmission line in terms of the phase field operator $\phi(x,t)=\int_{-\infty}^t E(x,t') dt'$, where  $E(x,t)$ is the electric field operator.  In the CPW, $\phi(x,t)$ is described by the massless, scalar Klein-Gordon equation in one dimension, the solution of which can be written as $\phi(x,t)= \phi_{\rm in}(x-c_0 t) +  \phi_{\rm out}(x+c_0 t)$, where $\phi_{\rm in(out)}$ is the field propagating inward to (outward from) the SQUID and $c_0\sim0.4c$ is the speed of light in the transmission line.  We solve the scattering problem in Fourier space defining 
\begin{displaymath}
\label{Eq:PhiTQuantizedDecomposed}
\phi_{\rm in(out)}
=
\sqrt{\frac{\hbar Z_0}{4\pi}}
\int_{0}^{\infty}\frac{d\omega}{\sqrt{\omega}}
\left(
  a_{\rm in(out)}(\omega) e^{-i(\mp k_\omega x + \omega t)} + h.c.
\right)
\end{displaymath}
where $a(\omega)$ and its hermitian conjugate $a^{\dagger}(\omega)$ and  are the standard annihilation and creation operators and $k_{\omega} = \omega/c_0$ is the wavenumber of the radiation. Solving the scattering problem then amounts to finding expressions for $a_{\rm out}^{\dagger}(\omega)$ and  $a_{\rm out}(\omega)$ as a function of $a_{\rm in}^{\dagger}(\omega)$ and  $a_{\rm in}(\omega)$. The boundary condition imposed by the SQUID determines the connection between these operators. With the output operators, we can then calculate the properties of the measurable output field assuming the input field is in a definite state, such as a thermal state or vacuum state.  For a static magnetic flux, $\Phi_{\rm ext}$, we obtain the simple expressions $a_{\rm out}(\omega) = R(\omega)a_{\rm in}(\omega)$ where $R(\omega)$ is the reflection coefficient from the SQUID. $R(\omega) = -\exp[2 i k_{\omega}\ell_{e}(\Phi_{\rm ext})]$ has the simple form\cite{Johansson:2009p2003} of a phase shift due to a transmission line of fixed length $\ell_{e}(\Phi_{\rm ext}) = L_{J}(\Phi_{\rm ext})/L_0$. Here $c_0= 1/ \sqrt{L_0 C_0}$, $L_0$ ($C_0$) is the inductance (capacitance) per unit length of the line, $L_J(\Phi_{\rm ext}) = (\Phi_0/2\pi)^2/E_J(\Phi_{\rm ext})$ is the Josephson inductance of the SQUID, $E_J$ is its Josephson energy, and $\Phi_0 = h/2e$ is the superconducting flux quantum.

In order to generate DCE radiation, $\ell_{e}$ must change with a nonuniform acceleration.  A simple example of this type of motion is just a sinusoidal drive with an amplitude of $\delta \ell_{e}$. If it is driven at $\omega_d$, we then find a simple expression\cite{Johansson:2010p3255} for $a_{\rm out}(\omega)$ in the region $\omega < \omega_d$:
\begin{equation}
a_{\rm out}(\omega) = R(\omega)a_{\rm in}(\omega) + S(\omega)a_{\rm in}^{\dagger}(\omega_d-\omega)
\end{equation}
where $S(\omega) =  -i(\delta \ell_{e} /c_0) \sqrt{\omega(\omega_d - \omega)}A(\omega)A^*(\omega_d - \omega)$ and $A(\omega)$ is the spectral amplitude of the transmission line.  Crucially, the time-dependent boundary leads to mixing of the input field's creation and annihilation operators.  With this expression we can calculate the output photon flux density for an input thermal state
\begin{equation}
\label{Output}
n_{\rm out}(\omega) = \langle a^{\dagger}_{\rm out}(\omega)a_{\rm out}(\omega) \rangle = n_{\rm in}(\omega) + |S(\omega)|^2n_{\rm in}(\omega_d-\omega)+ |S(\omega)|^2.
\end{equation}
The first two terms, proportional to $n_{\rm in}(\omega)$, represent the purely classical effects of reflection and upconversion of the input field to the drive frequency.  They are zero at zero temperature.  The last term is due to vacuum fluctuations and is, in fact, the DCE radiation.  

The photon production rate depends on the density of photonic states in the transmission line, which is $|A(\omega)|^2$.  For an ideal transmission line, $A(\omega)=1$ and the DCE radiation is $n_{\rm out}^{\rm DCE}(\omega) = (\delta \ell_{e} /c_0)^2\omega(\omega_d - \omega)$.  The integrated photon flux of the DCE radiation is $\Gamma_{\rm DCE} = (\omega_d/ 12\pi)(v_{e}/c_0)^2$ where $v_{e} = \delta \ell_{e}\omega_d$ is the maximum velocity of the boundary.  The relativistic nature of the effect is apparent here in that the photon flux goes to zero if we allow the speed of light to go to infinity.  Finally, we note that this spectrum is identical to that calculated for an ideal mirror oscillating in 1-D space\cite{Lambrecht:1996p1645}.  

\begin{figure}
\includegraphics[width = 0.5\columnwidth]{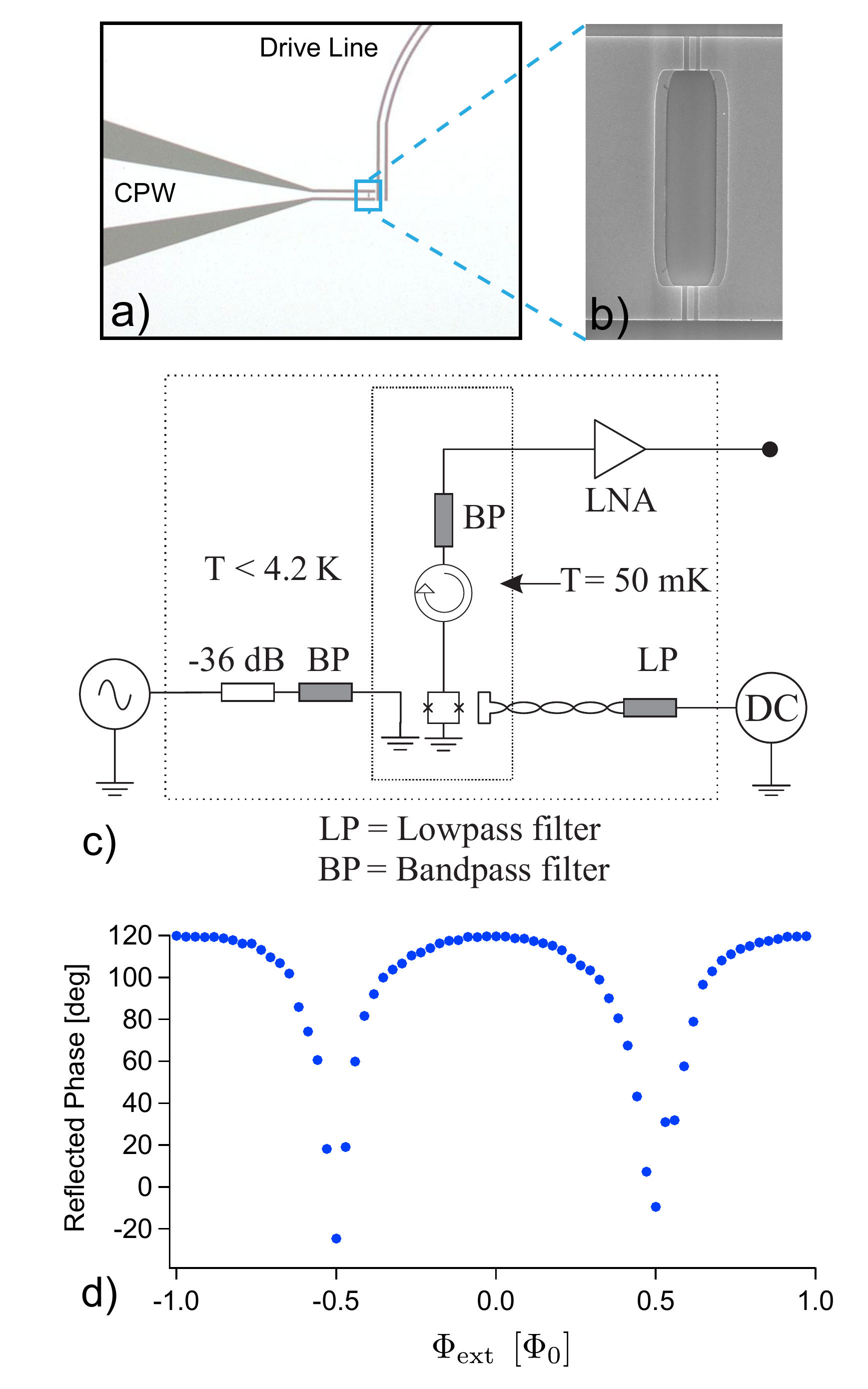}
\caption{a) Optical micrograph of the device. Light parts are Al while dark parts are the Si substrate.  The output line is labeled ``CPW" and the drive line enters from the top. Both lines converge near the SQUID. b) A scanning-electron micrograph of the SQUID.  The SQUID has a normal state resistance of 218 $\Omega$ implying a Josephson inductance at zero field of $L_J(0)=0.23$ nH. c) A simplified schematic of the measurement setup.  In addition to the driving line, a small external coil is used to apply a dc flux bias.  The driving line has 36 dB of cold attenuation along with an 8.4 to 12 GHz bandpass filter.  The filter ensures that no thermal radiation couples to the transmission line in the frequency region were we expect DCE radiation. The outgoing field of the CPW is coupled through two circulators to a cryogenic HEMT amplifier (LNA) with a system noise temperature of $T_N \sim 6 $ K.  The signal is further amplified at room temperature before being captured by two vector RF digitizers which use heterodyne mixing followed by digitization of the intermediate frequency signal. d) Measured phase shift of a microwave probe signal reflected from the SQUID, as it is tuned with a static flux. The probe frequency is 6.18 GHz.}
\label{Fig1}
\end{figure}

We present measurements on a sample with a short ($\sim 100\, \mu $m) Al coplanar waveguide (CPW) on-chip which transitions to a Cu CPW on a microwave circuit board (see Fig. 1a \& b). As a first measurement (see Fig. 1d), we can measure the phase shift of a microwave probe signal reflected from the SQUID as we change $\Phi_{\rm ext}$.  This illustrates how the SQUID changes the boundary condition at the end of the line.

We can now look at the effects of nonadiabatic perturbations.  The flux through the SQUID is driven at microwave frequencies by an inductively coupled CPW line that is short-circuited $\sim 20\, \mu $m from the SQUID.  The sample is cooled to 50 mK in a dilution refrigerator.  This corresponds to a thermal photon occupation number of $n = 0.008$ at 5 GHz, the center of our analysis band.  If we consider the last two terms in Eq (\ref{Output}) , which are the response of the system to the changing boundary, we can compare this small value $n = 0.008$ to the vacuum response, which has a coefficient of 1. We therefore conclude that thermal effects are negligible and disregard them in the rest of the paper.  For the first measurements, the drive is operated in continuous wave (CW) mode and the analysis frequency of the digitizer tracks the drive at $\omega_d/2$ with a 100 kHz bandwidth determined by digital filtering.  We study the total output power as a function of drive frequency and power.  The results are shown in Fig. 2a.  We clearly see photon generation for essentially all drive frequencies spanning the 8-12 GHz band set by the filtering of the line.  This corresponds to an analysis band of 4-6 GHz. The detailed drive power dependence of the output depends on the frequency and reflects a combination of the frequency response of both the drive line and the output line.

We can quantify the photon production rate by comparing it to the amplifier noise temperature. We see that the produced photons roughly double the noise level, suggesting a power per unit bandwidth of a few Kelvin.  We predict that in an ideal transmission line that the power should instead be a few mK.\cite{Johansson:2009p2003} In general though, transmission lines are not ideal and have parasitic resonances with a low quality factor ($Q$) associated with reflections at connectors, etc..  These resonances modify the electromagnetic density of states in the transmission line, $|A(\omega)|^2$, thereby enhancing\cite{Johansson:2010p3255} $\Gamma_{\rm DCE}$ by a factor of $\sim Q^2$.  In our measurement setup, we see such resonances, with $Q \sim\, $30-50, likely associated with the superconducting cable which runs between the sample and the amplifier.  This implies an enhancement of the photon production rate by a factor of 1000-2000, consistent with what we observe.

\begin{figure}
\includegraphics[width = 1\columnwidth]{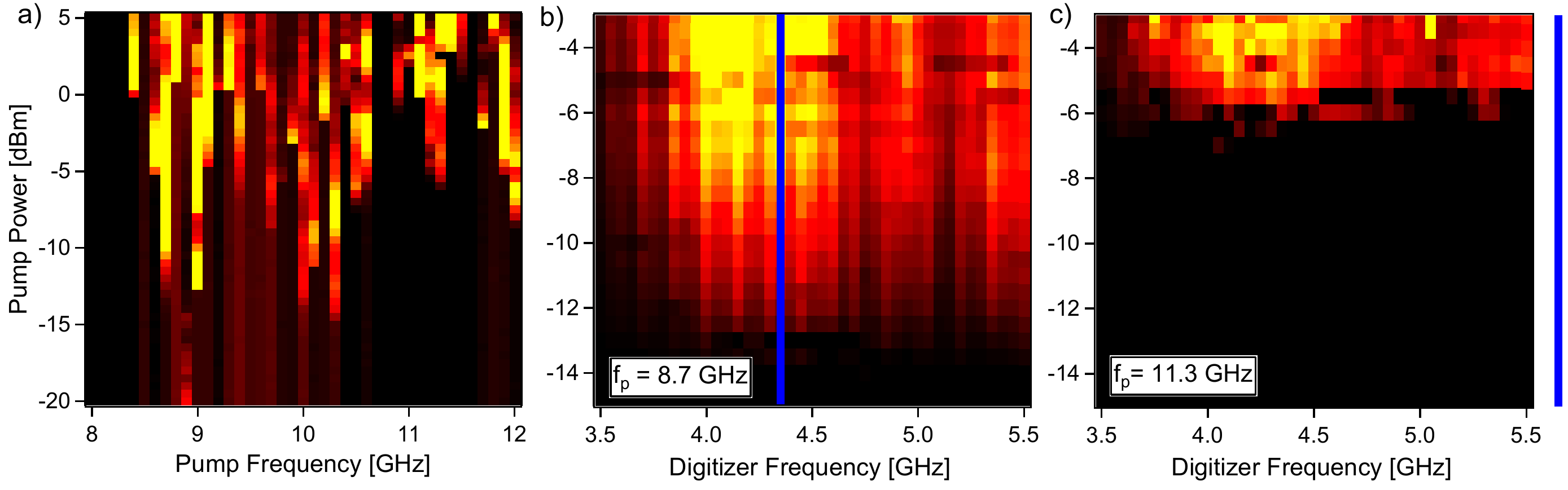}
\caption{Photons generated by the dynamical Casimir effect.  Here we show the output power of the transmission line while driving the SQUID.  Black is low power, while yellow is high power.  a) The drive frequency and power are scanned in continuous wave mode .  The digitizer tracks the drive frequency at $f_d/2$ with a 100 kHz analysis bandwidth.  b \& c) Broadband photon generation. The drive frequency is fixed while the digitizer frequency is scanned.  The drive is chopped on and off  and we record the difference in the power level for the on and off state.  The on (off) time is 50 ms (50 ms) and we measure in a 50 MHz analysis bandwidth. The drive frequencies in b) and c) are 8.70 GHz and 11.30 GHz respectively.  Half the drive frequency is indicated by the blue line. We see that for the two drive frequencies, the spectral shape of the output is similar. For instance, the output is more intense between 4-4.5 GHz in both cases.  This shows how the spectral density of the output line affects the radiation.   (The decline of the power below 4 GHz is due to the roll-off of the amplifier gain.) We also see that when half the drive frequency corresponds to the output mode, the necessary drive power decreases.
}
\label{Fig2}
\end{figure}

In the next set of measurements, we fix the drive frequency, but scan the digitizer frequency.  In this way, we can see over what band photons are produced for fixed drive frequency.  In Fig. 2 b \& c, we show the results for two different drive frequencies.  We clearly see broadband photon production at all analysis frequencies, including detunings from $f_d/2$ larger than 2 GHz. 

The broadband nature of the photon generation, both in terms of the drive frequency and analysis frequency, clearly distinguishes the observed phenomenon from that of a parametric amplifier comprised of a driven single-mode oscillator or cavity.  For a parametric amplifier, we expect the photon production to be narrow band in both drive and analysis frequencies, which is clearly not the case in this experiment.

Theory also predicts that the output should exhibit voltage-voltage correlations at different frequencies with a particular structure commonly known as two-mode squeezing (TMS).  Following Ref.\cite{CAVES:1985p2983}, we can describe a two-photon state, as we expect the DCE to generate, in terms of modulation of the center frequency of the state.  We can then define the modulation operators at the detuning $\epsilon$ as $\alpha_1(\epsilon) =(\lambda_+a(\omega_+) + \lambda_-a^{\dagger}(\omega_-))/\sqrt{2}$ and $\alpha_2(\epsilon) =(-i\lambda_+a(\omega_+) + i\lambda_-a^{\dagger}(\omega_-))/\sqrt{2}$ where $\omega_{\pm} = \omega_d/2 \pm \epsilon$ and $\lambda_{\pm} = (2\omega_{\pm}/\omega_d)^{1/2}$.  The factors $\lambda_{\pm}$ rescale the operators from quanta at $\omega_{\pm}$ to quanta at the center frequency $\omega_d/2$. We see that these operators mix excitations at the upper and lower sidebands of the field with a definite phase.  The TMS of the field then appears as an imbalance of the noise in one of these modes compared to the other.  We can then define a normalized squeezing statistic\cite{TMS}
\begin{equation}
\sigma_2 = \frac{\Sigma_{11}-\Sigma_{22}}{\Sigma_{11}+\Sigma_{22}}
\end{equation}
where $\Sigma_{mn} = \langle\alpha_m\alpha_n^{\dagger}+\alpha_n^{\dagger}\alpha_m\rangle/2$ is the symmetrized spectral density matrix.  For the case of an ideal transmission line,  we can calculate the TMS of the output field to be
\begin{equation}
\sigma_2 = 2\frac{\omega_+\omega_-}{\omega_d}\frac{\delta \ell_{e}}{c_0}\frac{1}{1+\omega_+\omega_-(\delta \ell_{e}/c_0)^2}.
\end{equation}
For small detunings ($\epsilon \ll \omega_d/2$), we get the approximate expression $\sigma_2 \approx v_{e}/2c_0$, which implies a squeezing of about 2\% for a 10\% modulation of the SQUID inductance.

Experimentally, we measure the four quadrature voltages of the upper and lower sidebands $I_{\pm}$ and $Q_{\pm}$.  The observable (hermitian) quadrature operators can be related to creation and annihilation operators as
\begin{equation}
I_{\pm} =\sqrt{\frac{\hbar\omega_{\pm}Z_0}{8\pi}}(a(\omega_{\pm}) + a(\omega_{\pm})^{\dagger}) \quad;\quad Q_{\pm} =-i\sqrt{\frac{\hbar\omega_{\pm}Z_0}{8\pi}}(a(\omega_{\pm}) - a(\omega_{\pm})^{\dagger}).
\end{equation}
We can then write $\sigma_2$ in terms of the quadratures as  
\begin{equation}
\label{sigIQ}
\sigma_2 = \frac{1}{P_{avg}}\left(\langle I_+ I_- \rangle - \langle Q_+ Q_- \rangle\right).
\end{equation}
where $P_{avg} = (\langle I_+^2 \rangle +\langle Q_+^2 \rangle+\langle I_-^2 \rangle +\langle Q_-^2 \rangle)/2$ is the average noise power in the sidebands. The theory also predicts a special structure for the correlations of the TMS, in particular that $\langle I_+ I_- \rangle = - \langle Q_+ Q_- \rangle$ and that $\langle I_+ Q_- \rangle = \langle I_- Q_+ \rangle$.  Finally, we comment\cite{CAVES:1985p2983} that $\Sigma$ transforms under phase rotations (see eqn. (\ref{psi})) such that we can specify $\langle I_+ Q_- \rangle = \langle I_- Q_+ \rangle = 0$ without loss of generality, which has been done in writing eq. (\ref{sigIQ}).


\begin{figure}
\includegraphics[width = 0.9\columnwidth]{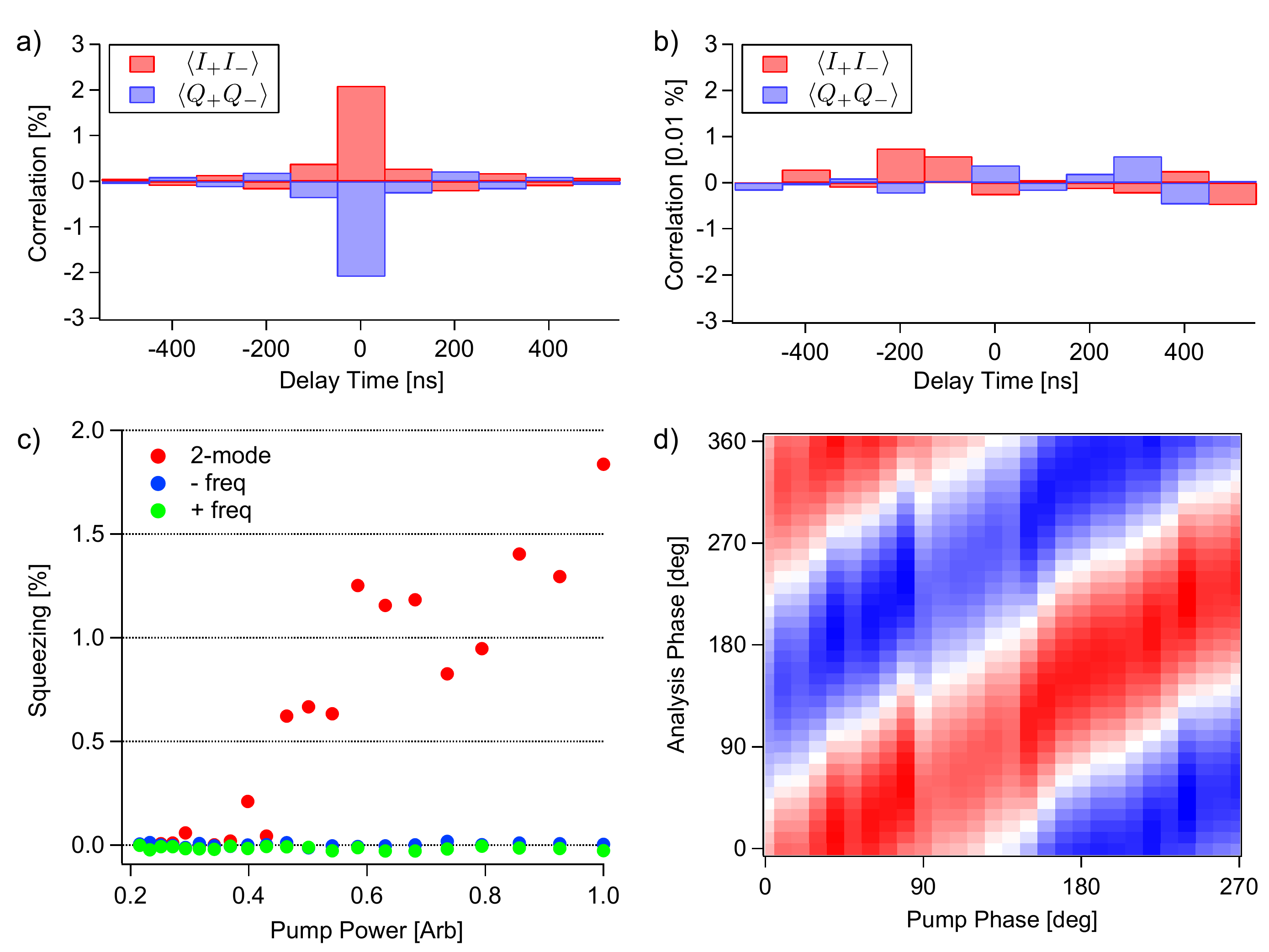}
\caption{Two-mode squeezing of the DCE field.  a) and b) The normalized cross-correlation functions $\langle I_+ I_- \rangle/P_{\rm avg}$ and $\langle Q_+ Q_- \rangle/P_{\rm avg}$ with the drive on (a) and drive off (b). Note that the scale of (b) is expanded by $\times100$.  The drive frequency is 11.30 GHz and the two digitizer frequencies are 5.63 GHz and 5.67 GHz.  The RF bandwidth is 10 MHz and $4\times10^7$ samples of each voltage are used.  With the drive on, (a), we clearly see cross correlations of a few percent.  Note that $\langle I_+ I_- \rangle = -\langle Q_+ Q_- \rangle$ as predicted. (The shape of the correlation functions in delay time is largely determined by the antialiasing filters of the digitizers.)  With the drive off, (b), we see that the residual correlations of the amplifier are negligible.  c) The two-mode squeezing, $\sigma_2$, of the field along with the one-mode squeezing, $\sigma_1$, at both $\omega_+$ and $\omega_-$ as a function of drive power.  Although there is some scatter in $\sigma_2$, we see that it clearly increases while the single-mode fields remain unsqueezed.  d) The effects of phase rotations on the complex correlation function $\Psi$ (defined in the text). The color scale is the real part of $\Psi$.  The $x$-axis corresponds to a change in the drive phase.  The $y$-axis corresponds to a digital rotation of the measured value of $\Psi$.  As expected for TMS, we see that $\Psi$ is symmetric under phase rotations.}
\label{Fig3}
\end{figure}

To measure the correlations, we use a single amplifier but take advantage of the fact that amplifier noise at different frequencies is uncorrelated.  After amplifying, we split the signal and feed the two outputs into two separate digitizers which are synchronized.  We then calculate the four $IQ$ cross-correlation functions.\cite{TCF}  Typical results are shown in Fig. 3.  In Fig. 3a \& b, we show the $\langle I_+ I_- \rangle$ and $\langle Q_+ Q_- \rangle$ cross correlations with the drive on and drive off for comparison.    With the drive on, we see very clear cross correlations which are $\sim100$ times larger than the parasitic amplifier correlation.  Also, we see that indeed $\langle I_+ I_- \rangle = - \langle Q_+ Q_- \rangle$ as we expect for two-mode squeezing.  We also see that this is not the case for the amplifier noise. The correlations imply a value of $\sigma_2 \sim 0.04$ consistent with our expectations.  The data in Fig. 3 is measured with a frequency seperation of 40 MHz ($\epsilon/2\pi = 20$ MHz), although we have measured similar values with separations as large as 700 MHz.


Theory predicts that even as the field is two-mode squeezed, if we look at either sideband frequency individually, we expect it to remain unsqueezed, essentially appearing as a thermal field at some effective temperature. In Fig. 3c, we then plot the TMS of the field, $\sigma_2$, along with the one-mode squeezing, $\sigma_1 = (\langle I^2\rangle -  \langle Q^2\rangle)/ (\langle I^2\rangle +  \langle Q^2\rangle)$ at both separate frequencies, $\omega_+$ and $\omega_-$, as a function of drive power.
Although there is some scatter in $\sigma_2$, we see that it clearly increases as a function of drive power while the one-mode fields remain unsqueezed.  This is an important additional check that the correlations arise from two-mode squeezing, and not just some well-timed modulation of, for instance, the amplifier gain that would produce the same type of apparent squeezing at the offset frequencies.

For a two-mode squeezed state, we can predict how the correlations transform under rotations of the phase of the EM field by an angle $\theta$.  In particular, if we define the appropriate combination of correlation functions
\begin{equation}
\Psi = (\langle I_+ I_- \rangle)-\langle Q_+ Q_- \rangle)+ i(\langle I_+ Q_- \rangle+\langle I_- Q_+ \rangle) 
\label{psi}
\end{equation}
we expect $\Psi$ to transform such that $\Psi' = e^{-2i\theta}\Psi$.  To explore this predicted symmetry, we can compute the complex quantity $\Psi$ from the experimental correlation functions and look at the rotation properties.  In Fig. 3d, we compare the results of rotating the measured quantity $\Psi$ to changing the drive phase.  We see, as expected for a TMS state, that the rotation of one cancels a rotation of the other. 

In conclusion, we observe broadband generation of microwave photons in an open transmission line with a periodically modulated boundary condition.  The emitted photons exhibit two-mode squeezing correlations, which are characteristic of photons generated in correlated pairs.  Taken together, we believe these results represent the first experimental observation of the dynamical Casimir effect.

We would also like to acknowledge G. Milburn and V. Shumeiko for fruitful discussions.  CW, PD, GJ, and AP acknowledge financial support from the Swedish Research Council, the Wallenberg foundation, STINT and the European Research Council.  FN and JJ acknowledge partial support from the LPS, NSA, ARO, DARPA, AFOSR, 
NSF grant No. 0726909, Grant-in-Aid for Scientific Research (S), MEXT Kakenhi on Quantum Cybernetics, and the JSPS-FIRST program.  TD acknowledges support from STINT and the Australian Research Council, grant numbers DP0986932 and FT100100025.

\bibliography{DCEpaper,Footnotes}

\begin{thebibliography}{22}
\bibitem[{TMS()}]{TMS}
\bibinfo{note}{Two-mode squeezing is often discussed in terms of the unitary
  squeezing operator $\Theta(r) = \exp[ra(\omega_+)a(\omega_-) -
  ra^{\dagger}(\omega_+)a^{\dagger}(\omega_-)]$ where $r$ is called the
  squeezing parameter. To connect to this language, one can show that $\sigma_2
  = -\lambda_+\lambda_-\tanh(2r)$. This then gives $r=(\delta
  \ell_e/c_0)\sqrt{\omega_+\omega_-}$.}
\bibitem[{TCF()}]{TCF}
\bibinfo{note}{In Fig. 3, we have displayed time correlation functions (TCF)
  which are the Fourier transforms of the frequency correlation functions (FCF)
  defined in the text. In particular, the value of the TCF at zero delay is the
  integral of the FCF in the measurement bandwidth. If we assume that the FCF
  is approximately constant in the measurement bandwidth, the integral reduces
  to multiplying the FCF by a constant factor. This is the same for all the
  TCF, and therefore cancels out of the normalized quantities displayed.}
\end{thebibliography}

\end{document}